# High On/Off Ratio Memristive Switching of Manganite-Cuprate Bilayer by Interfacial Magnetoelectricity

Xiao Shen[1,*,†], Timothy J. Pennycook[2,3,*], David Hernandez-Martin[4], Ana Pérez[4,*], Maria Varela[4,5], Yevgeniy S. Puzyrev[1], Carlos Leon[4], Zouhair Sefrioui[4], Jacobo Santamaria[4], and Sokrates T. Pantelides[1,5,6]

[1] Department of Physics and Astronomy, Vanderbilt University, Nashville, TN 37235

[2] SuperSTEM Laboratory, Daresbury, WA4 4AD, UK

[3] Department of Materials, Oxford University, Parks Road, Oxford OX1 3PH, UK

[4] Grupo de Fisica de Materiales Complejos, Universidad Complutense, 28040 Madrid, Spain

[5] Materials Science and Technology Division, Oak Ridge National Laboratory, Oak Ridge, Tennessee 37831, USA

[6] Department of Electrical Engineering and Computer Science, Vanderbilt University, Nashville, TN 37235

**Memristive switching serves as the basis for a new generation of electronic devices. Memristors are two-terminal devices in which the current is turned on and off by redistributing point defects, e.g., vacancies, which is difficult to control. Memristors based on alternative mechanisms have been explored, but achieving both the high On/Off ratio and the low switching energy desirable for use in electronics remains a challenge. Here we report memristive switching in a $La_{0.7}Ca_{0.3}MnO_3/PrBa_2Cu_3O_7$ bilayer with an On/Off ratio greater than $10^3$ and demonstrate that the phenomenon originates from a new type of interfacial magnetoelectricity. Using results from first-principles calculations, we show that an external electric-field induces subtle displacements of the interfacial Mn ions, which switches on/off an interfacial magnetic "dead" layer, resulting in memristive behavior for spin-polarized electron transport across the bilayer. The interfacial nature of the switching entails low energy cost about of a tenth of atto Joule for write/erase a "bit". Our results indicate**

---

[*] These authors contributed equally to this work.
[†] Correspondence to: xiao.shen@vanderbilt.edu,

**new opportunities for manganite/cuprate systems and other transition-metal-oxide junctions in memristive applications.**

The transistor, a three-terminal device, is the basic unit in modern electronics, which powers information technology. Driven by the approaching limits of transistor scaling, interest in exploring memristors as an alternative has recently surged. Memristors are devices in which the passage of current is controlled by exploiting the defining property of memristive materials, namely a pinched current-voltage hysteresis loop [1, 2]. In a transistor, the different conducting states are achieved by altering the gate voltage that externally modulates the carrier density in the semiconductor. In a memristor, one uses an external means to generate different conducting states by inducing reversible atomic displacements. Memristors often have the additional property of multiple conducting states of continuous tenability, which emulates biological synapses [3, 4].

Many materials have been explored for their memristive behavior. The most widely studied types are based on the dynamics of defects such as oxygen vacancies or metal ions [5, 6, 7]. Different resistance states are generated by either causing the formation and rupture of a conductive filament or modifying the Schottky barrier at the contacts. High On/Off ratios ($>10^3$) are often achieved, [8,9] but the defect-motion-based switching mechanism poses intrinsic challenges to the control of the materials, affecting the yield and reliability of the devices. Memristors based on other switching mechanisms have also been demonstrated, including molecular memristors [10] ferroelectric memristors [11], and spin-transfer torques (STTs) [12]. These memristors typically have On/Off ratio smaller than $10^2$ with the exception of ferroelectric memristors, which can reach $10^4$ [13], but switching entire ferroelectric domains entails high energy costs. Several recent papers demonstrated that the magnetic polarization in magnetic tunnel junctions (MTJs) [14] and magnetic-metal/ferroelectric junctions [15-17] can be modified by external electric fields, effectively producing memristive behavior, but the On/Off ratio is typically smaller than 10.

In this paper we report the discovery of memristive behavior with high On/Off ratio ($>10^3$) in transition-metal-oxide (TMO) interfaces. Such interfaces have been found to exhibit many other unusual properties. Here we focus on the well-known fact that at the interface of ferromagnetic (FM) and non-ferromagnetic TMOs such as La$_{0.67}$Ca$_{0.33}$MnO$_3$/YBa$_2$Cu$_3$O$_7$, there exists a "magnetic dead layer" (MDL) where the first layer of the FM TMO can be antiferromagnetically (AFM) coupled to the bulk of FM TMO [18]. We report a combined experimental and theoretical investigation demonstrating that an MDL can be turned on and off by an external electrical field in a La$_{0.7}$Ca$_{0.3}$MnO$_3$/PrBa$_2$Cu$_3$O$_7$ bilayer, causing memristive switching with On/Off ratio greater than 10$^3$. First-principles calculations show that the external electric field induces subtle displacements of the interfacial Mn atoms that control the presence or absence of an MDL, which then causes a memristive behavior for the transport of spin-polarized electrons. The high On/Off ratio and the subtle nature of the switching, which make the system very energy efficient (~0.1 atto Joule for write/erase a bit), make the LCMO/PBCO bilayer and similar manganite/cuprate systems particularly attractive for memristive devices.

La$_{0.7}$Ca$_{0.3}$MnO$_3$/PrBa$_2$Cu$_3$O$_7$ bilayers were grown using a high pressure (3.4 mbar) pure oxygen sputtering technique at elevated temperature (900º C), which is known to yield good epitaxial properties [19]. Standard optical lithography and ion milling from plasma source or electrically SiO$_2$ isolated mesas were used to define square micron size (4 x 4 um$^2$) pillars to measure perpendicular transport. The top electrode was evaporated silver. Transport (resistance versus field loops and I-V curves) was measured in a close cycle He cryostat equipped with an electromagnet that supplies a magnetic field up to 4000 Oe. For all measurements the top contact was grounded.

Aberration corrected scanning transmission electron microscopy (STEM) was employed to determine the atomic scale structure and composition of the LCMO/PBCO bilayer. High angle annular dark field (HAADF) imaging shows that the layers have grown coherently on the STO substrate as seen in Figure 1b. The atomic number contrast of HAADF imaging allows the STO, LCMO and PBCO layers to be easily identified. However because La, Pr and Ba are similarly heavy elements compared to the similarly light Cu and Mn elements the termination of the interface is not clear from the HAADF images alone. Atomic resolution spectrum imaging was therefore performed with electron energy loss spectroscopy. The maps show that the interface consists of a Mn-rich plane terminating the LCMO, adjacent to a single Ba-rich plane terminating the PBCO.

We therefore assume an $MnO_2$-BaO termination at the interface, as will be shown in our atomistic model below.

Electrical measurements in the current-perpendicular-to-plane geometry reveal a memristive hysteresis of the LCMO/PBCO bilayer, as shown in the current-voltage relation in Fig. 1c. Figure 1d shows the resistance hysteresis of the junction that was read with a voltage of 200 mV at 100 K. Under an external magnetic field of 4 kOe, the resistance (R) of the as-fabricated bilayer is around 300 Ω. By applying negative biases greater than -0.5 V, the measured resistance can be increased by more than 3 orders of magnitude, with larger maximum biases resulting in greater values of R. Applying a reverse positive bias switches the bilayer back to the initial low-resistance state. Such a large magnitude of resistance change is not observed in LCMO/PBCO/LCMO trilayer samples, which instead show a smaller factor of two change in resistance [20]. The resistance hysteresis varies with both temperature and magnetic field. The hysteresis window narrows as the temperature increases and vanishes at 180 K, which is above the Curie temperature of LCMO, as shown in Fig. 1e. This suggests that hysteresis is related to the spin-polarization of the current. Meanwhile, the hysteresis window widens as the magnetic field increases, with stronger magnetic fields yielding greater resistance at the high-resistance state (Fig. 1f).

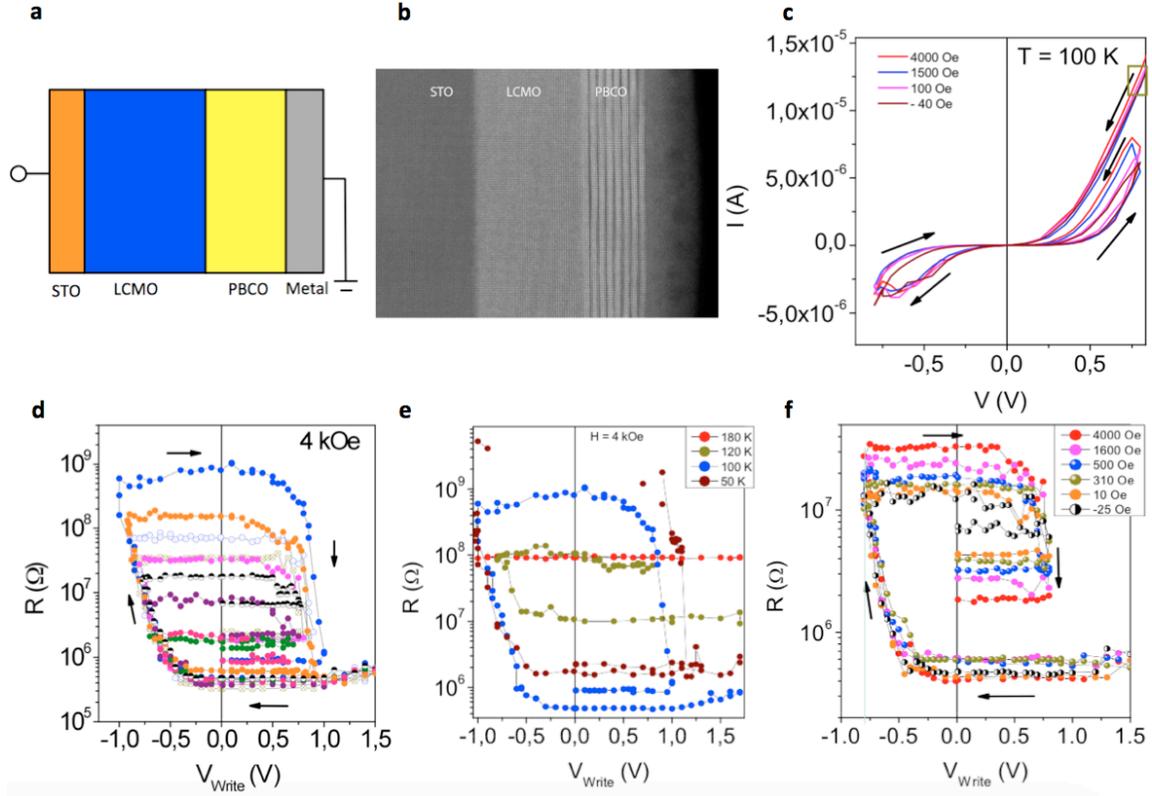

Figure 1. (a) The layout of the bilayer structure, where the top electrode (grounded) is shown in the right. (b) HAADF image of the LCMO/PBCO sample. (c) I-V hysteresis of the LCMO/PBCO bilayer recorded at 100 K. (d) Resistance hysteresis at 100 K. The different colors correspond to measurements with different maximum negative voltages. (e) Temperature dependence of the resistance hysteresis. (f) Magnetic field dependence of the resistance hysteresis.

The memristive behaviors of transition metal oxide films are often attributed to the diffusion of oxygen vacancies [4, 5, 7]. However, this mechanism is unlikely to be responsible for the observed switching here as hysteresis in resistance disappears when temperature is increased (above 180K) when vacancies are supposed to be more mobile. Moreover, we observe memristive hysteresis when the temperature is far below room temperature, meaning there is insufficient thermal energy to enable the vacancy motion despite the barrier being lowered by the external electric field [4, 5, 7, 21]. The diffusion barrier for oxygen vacancy migration in LCMO is 1.3 eV, [22] and the estimated maximum electric field during the experiment is 0.04 V/Å for a 1 V bias. For a hopping distance of 2 Å, the barrier lowering effect due to the electric field is 0.04 eV. For such an amount of barrier lowering, the diffusivity increases non-linearly with the electric field

[4, 21]. However the overall diffusivity is still too low for vacancy migration. Even if we use a generously overestimated value of 0.1 eV for the diffusion barrier lowering, the resulting diffusion barrier is still 1.2 eV. Although a diffusion barrier ~1 eV allows detectable vacancy motion at room temperature, [23] it is far too large for diffusion at 100 K, as a 1 eV barrier at 100 K is equivalent to a 3 eV barrier at 300K for diffusivity.

Other mechanisms for vacancy diffusion are also likely absent in the bilayers. In a good conductor with high current, electron wind can transfer energy to defects and cause their migration. However this mechanism is not in-play as the LCMO is a poor conductor and the PBCO barrier limits the current, which is further reduced when the bilayer is at the high resistance state. Recombination-enhanced diffusion [24, 25] can enable the diffusion at low temperature and also can cause memristive behavior of oxides, [26] but this effect is also unlikely present in the manganite/cuprate bilayers as there is no non-equilibrium concentration of electrons and holes and thus no carrier recombination. Besides the vacancy migration mechanism, in some ferromagnet/oxide/ferromagnet MTJs, spin-orbit coupling can rotate the direction of the magnetic moment of one of the ferromagnetic layers through voltage-controlled magnetic anisotropy (VCMA) [27] and cause memristive switching [14] with on/off ratio no greater than 10 [28]. This mechanism, however, does not explain the present observations, as only one magnetic layer is present in the bilayer samples. Therefore, a new mechanism must be at play.

It has been shown that at the perovskite manganite/cuprate interface, the magnetic moments on the interfacial Mn layer are significantly different from those in bulk manganite. At the LCMO/YBCO interface, a MDL can form [18] resulting from the interfacial Mn layer being coupled anti-ferromagnetically (AFM) to the ferromagnetic (FM) LCMO bulk (strongly suppressing double exchange transport through the interface). If such a MDL can be switched on and off in our manganite/cuprate bilayer by an electric field and the states are metastable (metastability in the magnetic states is known to exist in manganites [29]), the junction should also exhibit memristive hysteresis as illustrated in Fig. 2. In the low resistance state (LRS) the MDL is absent. The spin of the interfacial layer is FM coupled to the ferromagnetic LCMO bulk as in Fig. 2a, allowing the majority-spin electrons in the LCMO to tunnel through the PBCO, as shown in Fig. 2b (the light red area shows the additional barrier from the MDL). In this case, the current is large and R is small. In the high resistance state (HRS), the spin of the interfacial layer is AFM coupled with the LCMO bulk (Fig. 2c), giving rise to a MDL. This MDL adds an additional barrier to the tunneling of majority-spin electrons (Figure 2d), causing lower current and higher R.

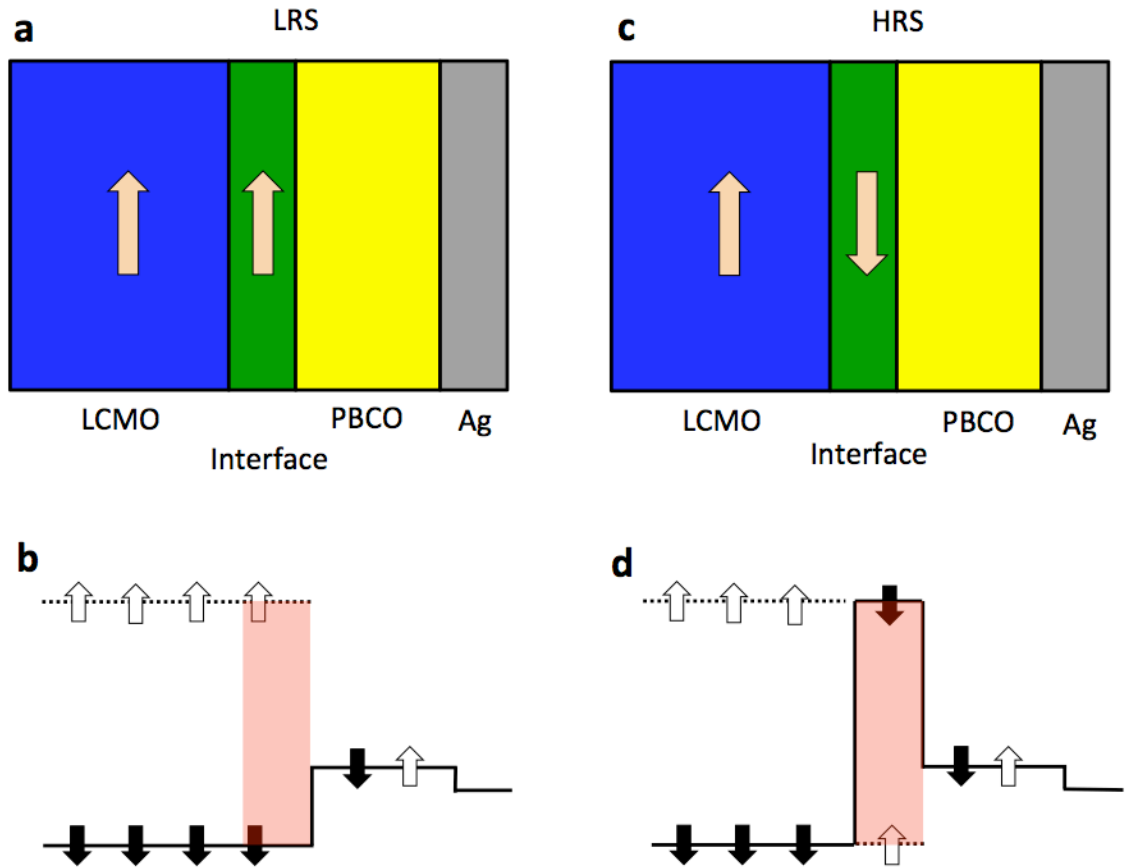

Figure 2. The mechanism of memristive switching at LCMO/PBCO interface. (a) The magnetic polarization at LRS, the arrow shows the direction of the magnetic moment. The potential "magnetic dead layer" is marked in green. (b) The band diagram of majority (black) and minority (white) spins at LRS. The light red area is the potential additional tunneling barrier if MDL is present. (c) The magnetic polarization at HRS. (d) The band diagram of majority and minority spins at HRS.

In Figure 3a, we show the relaxed structure of the LCMO/PBCO interface and highlight the position of the interfacial $MnO_2$ layer. It can be seen that the layer is polarized, with the Mn atoms displaced towards the PBCO. External biases, as illustrated in Fig. 3b,c, can alter the displacements of positively charged Mn atoms. As will be shown below, first-principles density functional theory (DFT) calculations confirm that such structural changes are indeed coupled to the change of magnetic ground state of the interfacial Mn layer and thus can switch the MDL on or off and cause the observed memristive hysteresis.

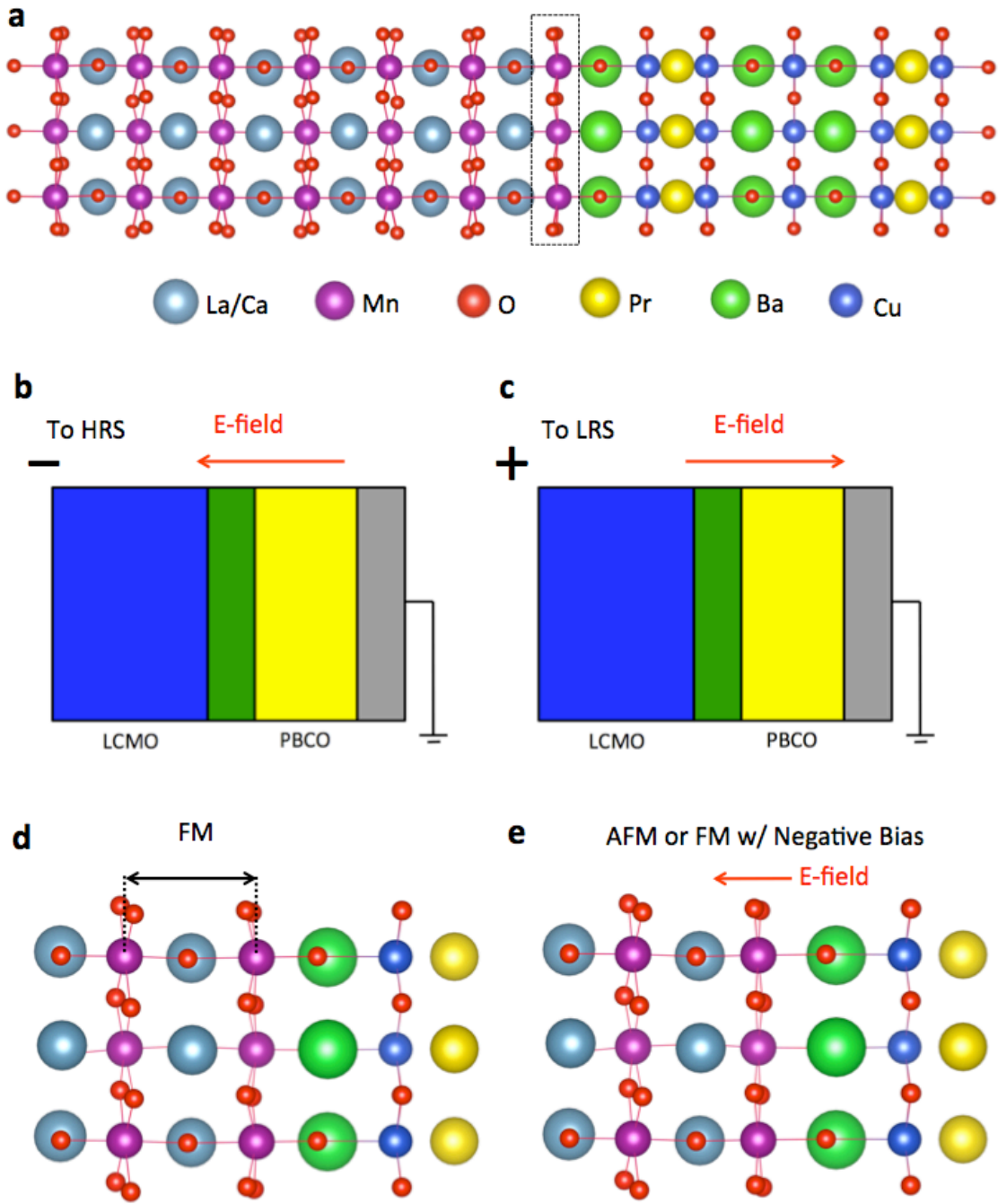

Figure 3. (a) The relaxed structure of the LCMO/PBCO interface. The position of possible "magnetic dead layer" is marked in dotted lines. (b, c) The polarities of the applied bias and the directions of the electric field inside the bilayer during the memristive switching. (d) The interface layer at optimized geometry when the interfacial Mn atoms are FM-coupled to bulk LCMO. This is also a zoomed-in view of Fig. 3a. The double-head arrow in black marks the distance between interface Mn layer the adjacent Mn layer as described in the text. (e) An exaggerated illustration of the interface layer at optimized geometry when the interfacial Mn atoms are AFM-coupled to bulk LCMO. It also illustrates the change of Mn displacement from (d) under a negative bias. The change of Mn displacement is magnified by 20 times to 0.2 A for demonstration. The actual change from DFT relaxation is only 0.01 Å as discussed in the text, which is not distinguishable from (d) to the eyes.

Using density functional theory, structural relaxations are performed for the PBCO/LCMO bilayer with fixed magnetic coupling between the interfacial Mn layer and the bulk in both the AFM and FM configurations. To capture the subtle structural difference, we use a very tight convergence setting that ensures the electronic self-consistency converges to $10^{-6}$ eV and the forces converge to $10^{-5}$ eV/Å. The relaxed structure with FM coupling has an energy 0.5 meV lower than the relaxed structure with AFM coupling, meaning the FM state is the ground state of the bilayer. Therefore, in the resting state, the MDL is not present and the bilayer is at LRS, as shown in Fig. 2 (a, b), consistent with electrical results shown in Fig. 1c.

The most important result from the calculations is that the displacements of interfacial Mn atoms are different in the cases of FM and AFM coupled states. In the FM coupled case, the distance between the interfacial Mn layer and the adjacent Mn layer (see Fig. 3d) is 3.946 Å; in the AFM coupled case, the value is 3.937 Å. Although the change of 0.01 Å from FM to AFM coupling seems quite small, it is significant when compared to the change of distances between bulk Mn layers from FM to AFM couplings, which do not exceed 0.002 Å. Multiple structural optimizations with different starting geometries are performed to ensure that these results are consistent and are not from artifacts of structural relaxation. Testing structural relaxations was also repeated at different doping levels and with different numerical setups (k-points, plane-wave cutoff) for the DFT calculations. The ~0.01 Å change in the distance between interfacial and adjacent bulk Mn layers that is much larger than the changes of distances between bulk Mn layers upon changing from FM to AFM coupled states are consistently obtained.

The DFT results on the displacements of interfacial Mn atoms are also consistent with the electrical data shown in Fig. 1c and the mechanism shown in Fig. 2. In the ground state, the Mn ions are displaced towards the PBCO by 3.946 Å from the adjacent Mn layer, as illustrated in Fig. 3d. In this state, the interfacial $MnO_2$ layer is FM coupled to the bulk LCMO, therefore MDL is absent and the bilayer is at LRS. Although the LCMO bulk in the bilayer is nominally metallic at 30% Ca doping [30], an electric field can be sustained at the interfacial $MnO_2$ layer because the Thomas-Fermi screening length corresponds to one to two units of LCMO [31]. Furthermore, as discussed below, the properties of the interface LCMO layer differ from bulk and might be viewed as an insulator. A negative bias (Fig. 3b) pulls the interfacial Mn towards the LMCO (Fig. 3e) and reduces the distance from the adjacent Mn layer. A change of 0.01 Å in the displacements of interfacial Mn layer preconditions the lattice structure to the optimized geometry of the AFM state. DFT calculations are performed to obtain the electronic ground states of the FM and AFM coupled states at the displaced configuration. The results show that the AFM-coupled state has an energy 0.3 meV lower than the FM coupled state. The magnetic moments on the interfacial Mn atoms relax into the AFM coupling state, activating the MDL and switching the bilayer to the HRS. A follow-up positive bias (Fig. 3c) pushes the interface Mn ions back towards the PBCO. Again, a 0.01 Å change in the displacement preconditions the structure to the optimized geometry of the FM state (Figure 3d). The magnetic moments on the interfacial Mn atoms then relax into the FM coupling state, deactivating the MDL and switching the bilayer back to the LRS.

As mentioned above, an 0.01 Å displacement of the interfacial Mn atoms "preconditions" the geometry of the lattice and facilitates the transitions between the magnetic states. To estimate the forces (and field) required to generate such displacements, we start from the relaxed structure of the FM-coupled state and displaced interface Mn ions towards LCMO by 0.01 Å. DFT calculations are performed for this geometry while keeping the magnetic configuration and other atoms frozen. The calculations show that the forces on the displaced Mn atoms are 0.25 eV/Å, which is the magnitude of the opposite force that needs to be generated by the applied electrical field.

Now we estimate the forces exerted on the interfacial Mn ions in the experiment and compare them with those calculated with DFT results. Depending on the level of Ca doping, $La_{1-x}Ca_xMnO_3$ can be a ferromagnetic insulator (x<0.2), a ferromagnetic metal (0.2<x<0.5), or an anti-ferromagnetic insulator (0.5<x) [30]. While the LCMO in our experiment has 30% Ca doping and therefore should be a ferromagnetic metal, the electron density around the Mn atoms at the interface is higher than that in the bulk, which is likely because the local bonding around the interfacial Mn atoms is slightly different from the bulk, as illustrated in Fig. 3a. For the interfacial Mn atoms, the calculated average electron density within the Wigner-Seitz cell is 11.574 $e$, which is greater than the value of 11.546 $e$ for the Mn atoms inside the LCMO part of the bilayer and is close to the calculated value of 11.579 $e$ for pure $LaMnO_3$. Therefore, we assume that the interfacial $MnO_2$ layer is similar to lightly doped $LaMnO_3$ and thus could be viewed as a ferromagnetic insulator. As a result, when an external bias is applied, the voltage drop is across both the interfacial $MnO_2$ layer and the PBCO layer. By assigning a thickness of 2 Å to the interface $MnO_2$ layer (half the spacing between Mn layers) and combining it with the thickness of the PBCO layer (~8nm) and the dielectric constants of lightly doped LMO ($\varepsilon$=18) [32] for the interfacial layer and PBCO ($\varepsilon$=80) [33], we estimate the electric field at the interface $MnO_2$ layer to be about 0.02 V/Å. The Born effective charge of interfacial Mn atoms from DFT calculation is 11. Combining these values together, we conclude that an external voltage of 0.5 V generates a force of 0.22 eV/Å on an interfacial Mn atom. Though this estimate is relatively crude, it is in good agreement with the DFT result that a force of 0.25 eV/Å is needed to displace the interfacial Mn atom by 0.01 Å and create a MDL.

Now we estimate the On/Off resistance ratio of the bilayer, which is the inverse of the transmission coefficient $T$ across the MDL. Using the WKB approximation for a rectangular barrier, $T$ equals $\exp(-2\sqrt{2m^*V_0/\hbar^2}d)$, where $m^*$ is the effective mass of electrons in LCMO, $V_0$ is the barrier height that equals the splitting between spin-up and spin-down electrons, and $d$ is the barrier width. Using $m^* = m_e$ [34], $V_0 = 3.3$ eV [35], and

$d = 3.9$ Å (MnO$_2$ layer thickness), we obtain $T = 7.5 \times 10^{-4}$ and an On/Off ratio of $1.3 \times 10^3$, in good agreement with the experimental results shown in Fig. 1d and 1e.

In addition to the high On/Off ratio, the mechanism described above is in good agreement with several other experimental observations. (1) The DFT calculations are independent of the electrical measurements and predict that negative bias switches the bilayer from the LRS to the HRS and that positive bias switches it from the HRS to the LRS, consistent with the experimental switching directions. (2) Stronger negative bias can cause a more complete formation of the MDL (convert larger portions of the interfacial plane to AFM domains) or make the MDL grow thicker, by propagating it beyond the first interfacial layer, thereby increasing the resistance in the HRS, as shown in experiments. (3) Meanwhile, once the MDL is destroyed, the bilayer has only one state, thus the resistance in the LRS is the same regardless of the applied positive bias, which is also observed experimentally. (4) The higher external magnetic field can enhance the degree of spin-polarization of injected current, that being "analyzed" by the AFM aligned interface plane in the MDL, therefore causing larger resistance in the HRS state, as the experiment shows. (5) The lack of a high On/Off ratio in LCMO/PBCO/LCMO trilayer can be understood as the trilayer has two opposing interfacial Mn layers that can be coupled by the magnetic moments in PBCO [20]. As the two interfacial Mn layers are opposing each other, they cannot be both switched from FM to AFM simultaneously by an electric field that produces displacements in the same direction. Therefore, the switching of one side is suppressed by its coupling to the opposite side through PBCO and a different mechanism may be at play in trilayers.

The physical phenomenon that underlies the memristive switching mechanism shown here is a new type of interfacial magnetoelectricity. It relies on the dependence of the magnetic coupling on the atomic positions, which is in accord with the fact that the magnetic properties in transition metal oxides can be very sensitive to the atomic structures [36]. It bears similarity to the recently observed interfacial magnetoelectricity in magnetic metal/ferroelectric oxide junctions [15-17] where the magnetism at the interface can be controlled by electrically reversing the polarization direction of the ferroelectric. First-principles calculations revealed that in those systems, the change of atomic displacements at the interface upon the switching of the ferroelectric plays the key role in altering the interfacial magnetism [37]. However, even though the magnetic switching in metal/ferroelectric systems only involves the interface, it is necessary to electrically switch the bulk of the ferroelectric. On the contrary, in the LCMO/PBCO bilayer, both magnetic switching and electrical switching are limited to the interface, which potentially offers faster switching and lower energy cost. Using the maximum force 0.25 eV/Å and the displacement of 0.01 Å, we estimated the energy cost to switching one Mn atom to be 1.3 meV. Therefore, the switching energy of a 10 nm by 10 nm area of the bilayer that representing a "bit" is only 0.8 eV, aka 0.13 atto Joule . In comparison, switching a $BiFeO_3$ ferroelectric memristor [13] with same area and a thickness of 4.6 nm would require 470 atto Joule (it takes 0.427 eV to switch one $BiFeO_3$ formula unit [38]); while for binary oxide TaOx memristors, the switching requires 100 fJ to 10pJ [39]. In addition, while the switching in metal/ferroelectric interfaces involves the change of the in-plane magnetic ordering that has a small effect on spin-current, the switching in LCMO/PBCO bilayer involves the activation of a magnetic "dead" layer, which allows high On/Off ratios for applications.

The key factors that enable the type of interfacial magnetoelectricity found in our system are the existence of a polarizable $MnO_2$ layer at the PBCO/LCMO interface and the modification of the magnetic properties of the Mn atoms by their displacement. As both factors are intrinsic to the perovskite manganite/cuprate interface, these findings should be applicable to other perovskite manganite/cuprate materials systems and possibly other transition metal oxides junctions as well. Further optimization of the materials might result in bilayer junctions with desired electrical properties to be used in a range of applications of memristive devices [4]. One key step would be realizing devices that operate at room temperature. This may be achievable by replacing $LaCaMnO_3$ with $LaSrMnO_3$, whose Curie temperature can be higher than 300 K [40].

**METHODS**

**Sample growth.** The samples were grown on top of STO (001) substrates using a high pressure (3.2 mbar) pure oxygen sputtering deposition system at high temperature (900 °C) [19, 20]. We fabricated magnetic tunnel junctions from [PBCO (8 nm)/LSMO 50 nm] bilayers using standard UV optical lithography and ion milling. We patterned samples into micron size (9x18 μm$^2$ and 5x10 μm$^2$) rectangle shape pillars and measured their magnetotransport properties. For transport properties we deposited Ag top contacts on the PBCO. Typically 40% of junctions per sample were not shunted and could be measured, which represents a large success ratio of our patterning process. IV curves were measured using current source and voltmeter. For all measurements the top contact was grounded such that negative (positive) voltages correspond to electric fields pointing downwards (upwards).

**Scanning transmission electron microscopy.** High angle annular dark field (HAADF) imaging was performed on Nion UltraSTEM 100 and UltraSTEM 200 instruments operated at 100 and 200 kV respectively. Both microscopes use cold field emission electron sources and aberration correctors capable of neutralizing up to fifth order aberrations. Electron energy loss spectroscopy was performed on the Nion UltraSTEM 100 using a Gatan Enfina EEL spectrometer.

**Density functional theory calculations.** We employ the Perdew–Burke–Ernzerhof (PBE) [41] version of the generalized-gradient approximation (GGA) exchange-correlation functional. We use projector augmented wave potentials [42] and plane-wave basis as implemented in the Vienna Ab-initio Simulation Package (VASP) code. [43] The kinetic energy cutoff of the plane wave basis is set to be 368.6 eV. The electronic self-consistent calculations are converged to $10^{-5}$ eV between two self-consistent steps. The structural relaxations are converged to $10^{-4}$ eV for the total energy difference between two ionic steps. The simulation cell consists of a layer of $La_{0.67}C_{0.33}MnO_3$ of 23 Å and a $PrBa_2Cu_3O_7$ layer of 19 Å, with a total of 104 atoms. Brillouin zone sampling is performed by using the 2x2x1 k-point-mesh. To account for the electron correlations, we use an implementation [44] of DFT+U methods [45] and apply U=2 eV on Mn 3d orbitals. The 4f electrons on Pr atoms are kept frozen at the core.

## ACKNOWLEDGMENTS

The work at Vanderbilt was supported by National Science Foundation grant DMR-1207241, by Department of Energy grant DE-FG02-09ER46554, and by the McMinn

Endowment at Vanderbilt University. Computational support was provided by the NSF XSEDE under Grant # TG-DMR130121. Research at UCM was supported by Spanish MICINN through grants MAT2011-27470-C02 and Consolider Ingenio 2010-CSD2009-00013 (Imagine), by CAM through grant S2014/MAT-PHAMA II. Research at SuperSTEM, the UK National Facility for Aberration-Corrected STEM was supported by the EPSRC.